\documentclass[aps,amsmath,twocolumn,preprintnumbers,floatfix,nofootinbib,raggedfooter,twoside,showpacs,showkeys,a4paper]{revtex4}
\usepackage{ifpdf}\ifpdf\pdfoutput=1\fi  

\usepackage{psfrag,graphicx}
\makeatletter\def\input@path{{../origfigs/}}\makeatother
\graphicspath{{../origfigs/}}

\PassOptionsToPackage{notightpage}{pst-pdf}

\ifpdf
  \usepackage{pst-pdf}
  \renewcommand{\PDFcontainer}{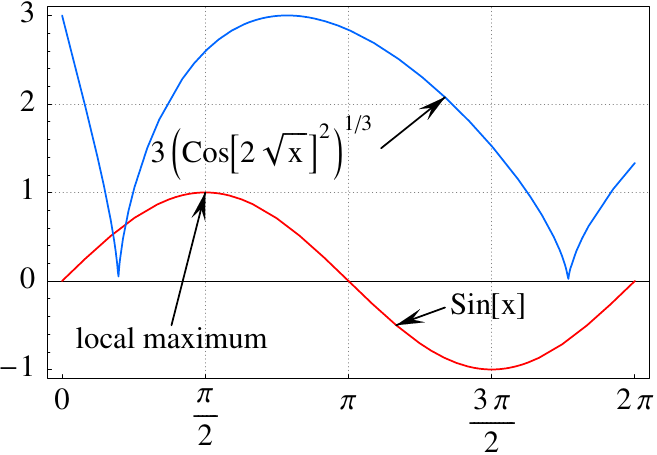}
\fi
\usepackage{xcolor}

\usepackage{fancyvrb}
  \DefineVerbatimEnvironment{MyVerbatim}{Verbatim}{frame=lines,fontsize=\small,framesep=5pt,samepage=true}
\usepackage{subfigure}
\usepackage[T1]{fontenc} 
\usepackage{lmodern}
\usepackage{textcomp}             
\usepackage[
  breaklinks,  
  urlbordercolor={0.9 0.5 0.5},
  citebordercolor={0.5 0.9 0.5},
  linkbordercolor={0.7 0.7 1},
  hyperfootnotes = false]{hyperref}   
\ifpdf\else\usepackage{breakurl}\fi 
\newcommand{\pack}[1]{\NoCaseChange{\textsf{#1}}}     
\newcommand{\fname}[1]{\texttt{#1}}                   
\newcommand{\mc}[1]{{\texttt{#1}}}                    
\newcommand{\acro}[1]{{\small #1}}                    
\providecommand{\cs}[1]{%
   \texttt{\expandafter\string\csname #1\endcsname}}  
\providecommand{\latex}[1]{\texttt{#1}}               
\providecommand{\env}[1]{\latex{\{#1\}}}              
\providecommand{\marg}[1]{$\langle$\texttt{\textsl{#1}}$\rangle$} 
\providecommand{\oarg}[1]{$[$\texttt{\textsl{#1}}$]$} 
\newcommand{\Rule}{\(\to\)\allowbreak}                
\newcommand{\amsmath}{\texttt{amsmath}}              
\newcommand{\Mathematica}{\NoCaseChange{\textsc{Mathematica}}}      
\newcommand{\MathematicaR}{\Mathematica\textsuperscript{\tiny\textregistered}}
\newcommand{\PSfrag}{\pack{PSfrag}}
\newcommand{\pstpdf}{\pack{pst-pdf}}
\newcommand{\Ghostscript}{\pack{Ghostscript}}
\newcommand{\dvips}{\fname{dvips}}
\newcommand{\PostScript}{\textsc{Post\-Script}}
\newcommand{\PDFLaTeX}{pdf\LaTeX}
\newcommand{\MathPSfrag}{\pack{MathPSfrag}}
\newcommand{\MacOSX}{MacOS~X}
\newbox\subfigbox
\makeatletter
\newenvironment{subfig}{%
  \def\caption##1{\gdef\subcapsave{\relax##1}}%
  \let\subcapsave\@empty
  \setbox\subfigbox\hbox
  \bgroup}
  {\egroup
  \subfigure[\subcapsave]{\box\subfigbox}}
\makeatother

\begin{document}
\title{\MathPSfrag\ 2: Convenient \LaTeX\ Labels in Mathematica}
\author{Johannes  Gro\ss{}e$^*$}
\affiliation{Institute of Physics, Jagiellonian University, Reymonta 4, 30-059 Krak\'{o}w, Poland}
\date{January 15, 2008}

\begin{abstract}
This article introduces the next version of \MathPSfrag. \MathPSfrag\ 
is a \MathematicaR\ package that during export \emph{automatically}
replaces all expressions in a plot by corresponding \LaTeX\ commands.
The new version can also produce \LaTeX\ independent images; e.g.,
\acro{PDF} files for inclusion in \PDFLaTeX. Moreover from these
files a preview is generated and shown within \Mathematica.
\end{abstract}

\maketitle


\section*{Introduction}
Many programs producing \acro{EPS} graphics do not allow the inclusion
of \LaTeX\ commands. While there exist several solutions to work
around these difficulties, they all have various drawbacks. (See
\cite{McKay:1999} for a discussion of several methods and an
alternative approach to overcome these difficulties.)  In this
article, we will focus on a particular existing solution, the
\PSfrag{} package \cite{Grant:1998}, which provides \LaTeX\ macros
allowing the user to replace pieces of text (``tags'') in an \acro{EPS} file by
an arbitrary \LaTeX\ construct.

However, for \PSfrag{} to work, the application must write tags
unaltered into the \acro{EPS} file. For \Mathematica{}
\cite{Wolfram:1999book,Wolfram:2005mathematica}, this requirement
amounts to using single words, strictly consisting of alphanumeric
characters only. As a consequence, the user has to work most of the
time with an inconveniently labeled graphic and is furthermore
required to keep track of the tags used in the substitution macros.

On the other hand, it is not always possible to use \Mathematica{}'s
conventional export function as it produces \acro{EPS} files requiring
the inclusion of additional fonts into the document. This means
configuring the local \TeX\ installation such that it finds the fonts
provided by Wolfram Inc.\ \cite{WRI:tetex,WRI:gsfont,WRI:intoeps}, a
process often not being under the author's control. A way out is to
include the fonts into the \acro{EPS} file and set the font family to
a standard \PostScript\ one. The required steps are version dependent---
the latest version as of this writing, 6.0, has sane defaults---for 5.2, 
the following commands are required:
\begin{MyVerbatim}[commandchars=\\()]
Plot[\dots, TextStyle\Rule{FontFamily\Rule"Times"}]
Export[\dots, ConversionOptions\Rule
\hfill{"IncludeSpecialFonts"\Rule()True}]
\end{MyVerbatim}
However, automatic inclusion of \Mathematica's special fonts, which
irrespective of the chosen \mc{FontFamily} are used for displaying
important symbols like brackets, is only a last resort.  While the
slight mismatch between a standard \PostScript\ font's appearance,
(here Times Roman, cf.~fig.~\ref{fig:poor}), and that of \LaTeX's
standard font (Computer Modern) may be acceptable in case of ordinary
text labels, mathematical expressions like square roots or fractions
cannot compete with \LaTeX's typesetting quality in this approach.

\begin{figure}[t]
\vspace{4.5ex}
\begin{subfig}
  \includegraphics[width=75mm]{ex_nopsfrag.eps}
  \caption{Conventional \Mathematica{} plot \emph{before} using \MathPSfrag\label{fig:poor}}
\end{subfig}
\begin{subfig}
  \ifpdf\else\input{ex_auto-psfrag.tex}\fi  
  \mbox{\includegraphics[width=75mm]{ex_auto-psfrag.eps}}

  \caption{The same plot \emph{after} automatically substituting all \mc{Text} 
  primitives (including tick mark labels) by \LaTeX\ output.\label{fig:beautiful}}
\end{subfig}
\caption{Old vs.\ new graphics export mechanism.}
\end{figure}
\begin{figure}[b]
\vspace{1.5ex} 
\footnoterule\footnotesize
\vspace{-2ex} 
\raggedright
\(\sp*\) \texttt{jgrosse@th.if.uj.edu.pl}\\
\(\phantom{\sp*}\)   \url{http://wwwth.mppmu.mpg.de/members/jgrosse/mathpsfrag}
\vspace{-0.5ex}
\end{figure}

\begin{figure*}
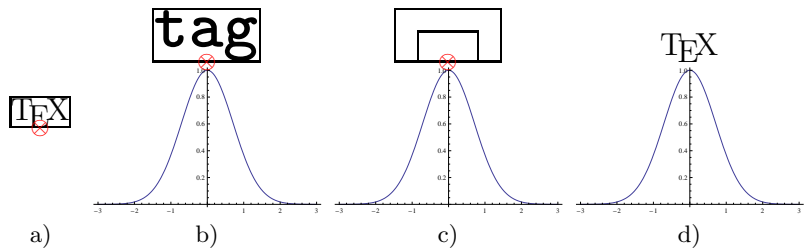

\begin{tabular}{cccc}
\raisebox{10mm}{
\begin{psfrags}
     \psfrag{gA}[bc][bc]{\fboxsep0pt\fbox{\large\TeX}}
     \psfrag*{gA}[cc][bc]{\textcolor{red}{$\otimes$}}
     \includegraphics[width=15mm]{testfig.eps}
\end{psfrags}
}
&
\begin{psfrags}
     \psfrag{tag}[bc][bc]{\fboxsep0pt\fbox{\Huge\textbf{\texttt{tag}}}}
     \psfrag*{tag}[cc][bc]{\textcolor{red}{$\otimes$}}
     \includegraphics[width=30mm]{tagplot-psfrag.eps}
\end{psfrags}
&
\begin{psfrags}
     \psfrag{tag}[bc][bc]{\fboxsep0pt\fbox{\Huge\textbf{\texttt{\phantom{tag}}}}}
     \psfrag*{tag}[bc][bc]{\fboxsep0pt\fbox{\phantom{\large\TeX}}}
     \psfrag*{tag}[cc][bc]{\textcolor{red}{$\otimes$}}
     \includegraphics[width=30mm]{tagplot-psfrag.eps}
\end{psfrags}
&
\begin{psfrags}
     \psfrag*{tag}[bc][bc]{\large\TeX}
     \includegraphics[width=30mm]{tagplot-psfrag.eps}
\end{psfrags}\\
a)&b)&c)&d)
\end{tabular}
\caption{Action of \texttt{\cs{psfrag}\{tag\}[bc][bc]\{\cs{TeX}\}}.
  The first coordinate pair \texttt{[bc]} picks the reference point (red
  crossed circle) in the \TeX\ expression (fig.~a), the second one in
  the \texttt{tag} that is part of the original \acro{EPS} file
  (fig.~b).  Positioning is achieved by overlaying both boxes such
  that their respective reference points coincide (fig.~c). The final result is shown
  in fig.~d.  Scales are exaggerated for better illustration.
  \label{fig:tagplot}}
\end{figure*}

Font inclusion is also a feature that---like the visually displeasing
option to not use special fonts at all---has become available only
starting from \Mathematica\ version~4.2.1. Consequently, some authors
simply restrict labeling of \Mathematica{} plots to a bare minimum.

\MathPSfrag{} \cite{Grosse:2005} is a package that conveniently
produces publication-quality labels in \acro{EPS} files generated by
\Mathematica{}.  \MathPSfrag{} automates many (often all) tedious
details related to the use of the standard \LaTeX\ package \PSfrag{},
while still allowing manual fine tuning. As a demonstration of the
degree of automation, compare fig.~\ref{fig:poor}, which has been
generated by using the standard \Mathematica{} command \mc{Export},
and fig.~\ref{fig:beautiful}, generated by \MathPSfrag's export
instruction.

While the solution presented here, relies on the \PSfrag{} package, it
avoids many of its shortcomings by providing a semi-automatic layer.
In many cases it is sufficient to simply use the new \mc{PSfragExport}
command.

\MathPSfrag\ also allows the creation of stand-alone images that do
not need any additional preparations in the manuscript.  This is
achieved by calling \LaTeX\ and \Ghostscript\ from within
\Mathematica\ requiring the user to have a full \LaTeX\ distribution
on the same machine where \Mathematica\ resides.

The first section of this document describes the \PSfrag\ \LaTeX\ 
package, while the second and third section explain its counterpart \MathPSfrag.
A number of small examples are shown followed by a discussion of
how to produce \acro{PDF} files. Before the conclusions, a few
limitations of the current approach and its implementation are discussed.

\section{PSfrag\label{sec:psfragsty}}

This is intended to be a short introduction to the \LaTeX\ package
\PSfrag{} explaining only the essential features necessary to
understand the corresponding \Mathematica{} package's internals and to
take advantage of its manual options if automatic placement does not
yield the desired result. The full documentation can be found in
\cite{Grant:1998}. 

\PSfrag{} provides the macro
\begin{MyVerbatim}[commandchars=\\()]
\cs(psfrag){\marg(tag)}[\oarg(texposition)][\oarg(psposition)]
\hfill[\oarg(scale)][\oarg(rot)]{\marg(\textrm(\LaTeX))}
\end{MyVerbatim}
which replaces any occurrence of \marg{tag} in the output of an
\acro{EPS} file by \marg{\textrm{\LaTeX}}.  According to
\cite{Grant:1998}, ``all \cs{psfrag} calls that precede an
\cs{includegraphics} (or equivalent) in the same or surrounding
environments'' will affect the output of the included graphics; i.e.,
\cs{psfrag} commands can be defined either locally, to act on strictly
one graphic, or globally, thus acting on all graphics in a document.
The mechanism is demonstrated in fig.~\ref{fig:tagplot}.
\begin{figure}
     \psfragdebugon
     \psfrag{gA}[br][br]{---[br][br]---}
     \psfrag*{gA}[Br][bc]{---[Br][bc]---}
     \psfrag*{gA}[cr][bl]{---[cr][bl]---}
     \psfrag*{gA}[tr][Bl]{---[tr][Bl]---}
     \psfrag*{gA}[bc][Bc]{---[bc][Bc]---}
     \psfrag*{gA}[Bc][Br]{---[Bc][Br]---}
     \psfrag*{gA}[cc][cr]{---[cc][cr]---}
     \psfrag*{gA}[tc][cc][0.75][45]{---[tc][cc][0.75][45]---}
     \psfrag*{gA}[bl][cl][1.5][30]{---[bl][cl][1.5][30]---}
     \psfrag*{gA}[Bl][tl]{---[Bl][tl]---}
     \psfrag*{gA}[bl][Bl]{~~~~~(baseline)}
     \psfrag*{gA}[bl][cl]{~~~~~(center line)}
     \psfrag*{gA}[bl][tc][1][-90]{~~~~~(center line)}
     \psfrag*{gA}[cl][tc]{---[cl][tc]---}
     \psfrag*{gA}[tl][tr][1][180]{---[tl][tr][1][180]---}
     \includegraphics[width=55mm]{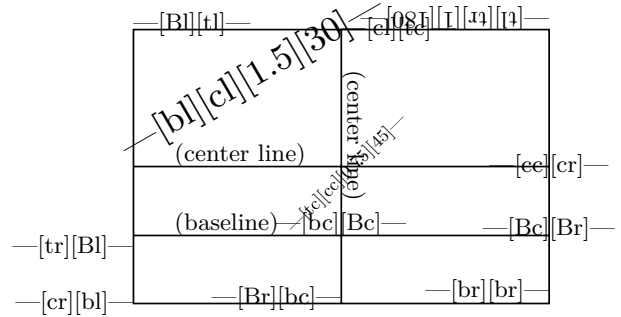}
\caption{Illustration of the various optional arguments of the
\cs{psfrag} command, taken from \protect\cite{Grant:1998} with minor changes.
The first option determines the alignment of the \LaTeX\ description,
while the second one is responsible for the point to which the \LaTeX\
macro is attached.
\label{fig:psfragopts}}
\end{figure}

\oarg{texposition} and \oarg{psposition} are optional arguments that
allow setting (first) the vertical (top, bottom, Baseline, or center)
and (second) the horizontal (left, right, center) alignment of the
replacement text by specifying the respective first character of the
choices given in parentheses.  The arguments refer to the position of
the reference point in the respective bounding boxes.  The \LaTeX\
construct is placed such that its reference point is at the position
of the corresponding \PostScript\ (tag) box' reference point,
cf.~fig.~\ref{fig:psfragopts}.

\oarg{scale} and \oarg{rot} permit scaling and rotation of the
inserted box, where the rotation (in degree) is relative to the slope
of the \PostScript\ bounding box such that a value of ``0'' preserves
the orientation, see fig.~\ref{fig:rot}.  Scaling is best achieved by
using \LaTeX\ scaling commands, like \cs{Large}, instead of the
\oarg{scale} option, since the standard \LaTeX\ fonts consists of
bitmaps rendered specifically for the chosen size and do not rescale
well.

Since \PSfrag\ exchanges the labels of an \acro{EPS} image, these may
protrude from the bounding box of the resulting image. Fortunately,
\cs{includegraphics} offers the options \latex{bb} and \latex{trim},
which may be used to override or correct the obsoleted information in
the \acro{EPS} file. In particular, in conjunction with some of the
\acro{PDF} production techniques described in subsequent sections,
correct(ed) bounding boxes can be important because protruding
material might be clipped.

\section{MathPSfrag}

There are only three commands needed to control \MathPSfrag's
\acro{EPS} generation: \mc{PSfragExport}, which supersedes
\Mathematica's \mc{Export} command, and \mc{PSfrag}, which allows
overriding of the defaults for particular expressions.  In addition
the \mc{UnPSfrag} command is provided which calls \LaTeX\ and
\Ghostscript\ to carry out the \PSfrag\ replacements and create an
(ordinary) \acro{EPS} and \acro{PDF} image that can be included with
the usual \cs{includegraphics} command.

The export function is usually called in conjunction with 
\mc{UnPSfrag}:
\begin{MyVerbatim}[commandchars=\\()]
PSfragExport[\marg(basename), \marg(graphics), \oarg(options)]
  // UnPSfrag;
\end{MyVerbatim}
This converts \marg{graphics}, the usual \mc{Graphics} construct
returned by \Mathematica{} commands like \mc{Plot}, to an \acro{EPS}
file and a \LaTeX\ file containing \cs{psfrag} macros. The
\mc{UnPSfrag} command, if provided, will merge these two into a single
\acro{EPS} file and will also produce a \acro{PDF} version. More about
this in section \ref{sec:unpsfrag}.

There are a number of options for \mc{PSfragExport} that can be used
to override internal assumptions about the automatic processing of
graphics. They are documented in the manual accompanying \MathPSfrag.
The respective file names of the \LaTeX\ and \acro{EPS} file are
determined by \marg{basename} to which the value of the options
\mc{TeXSuffix} and \mc{EpsSuffix} is appended.
\begin{itemize}
  \item \mc{TeXSuffix\Rule{}"\marg{string}"}\hfill(\mc{"-psfrag.tex"})
  \item \mc{EpsSuffix\Rule{}"\marg{string}"}\hfill(\mc{"-psfrag.eps"})
\end{itemize}
Unless a base name containing a path is given, the output is written
to the current directory, which can be set using \Mathematica's
\mc{SetDirectory}.  Unknown options are passed on to the graphics
using a \mc{Show} command.

It may happen that the user is not satisfied with the automatic output
generated by \mc{PSfragExport}. Therefore it is possible to manually
control the replacement of any expression by wrapping
\begin{MyVerbatim}[commandchars=\\()]
PSfrag[\marg(expr), \oarg(options)]
\end{MyVerbatim}
directly around any \Mathematica{} expression \marg{expr} appearing as
text in a graphic, such as the argument of a \mc{PlotLabel\Rule...} or
\mc{AxesLabel\Rule...} option or a \mc{Text} graphics directive. As a
simple example, consider
\begin{MyVerbatim}[commandchars=\\()]
p=Plot[\dots,PlotLabel\Rule()PSfrag["\(\chi\)^2-test",
   TeXCommand\Rule()"$\textbackslash\textbackslash()chi^2$-test"]];
PSfragExport["chisquare", p];
\end{MyVerbatim}
This is also the simplest possibility to correct malformed
automatically generated \LaTeX\ code.

\mc{PSfrag} processes the following options, whose defaults have
been put in parentheses.
\begin{itemize}
  \item \mc{TeXCommand\Rule{}"\marg{string}"}\hfill(\mc{Automatic})
  \item \mc{PSfragTag\Rule{}"\marg{string}"}\hfill(\mc{Automatic})
  \item \mc{TeXPosition\Rule{}"\marg{yx}"}\hfill(\mc{Automatic})
  \item \mc{PSPosition\Rule{}"\marg{yx}"}\hfill(\mc{CopyTeXPosition})
  \item \mc{PSRotation\Rule{}\marg{number}}\hfill(0)
  \item \mc{PSScaling\Rule{}\marg{number}}\hfill(1)
  \item \mc{TeXShiftX\Rule{}"\marg{texdim}"}\hfill(0pt)
  \item \mc{TeXShiftY\Rule{}"\marg{texdim}"}\hfill(0pt)
\end{itemize}

Actually, \mc{PSfragExport}'s automatic mechanism basically wraps
\mc{PSfrag} around all \mc{Text} primitives using the default values
above.  However, manual wrapping has the advantage of allowing
different options to be applied to expressions where the automatic
behavior did not give satisfactory results.

\mc{TeXCommand\Rule{}"\marg{string}"} uses \marg{string} as the
\LaTeX\ command to appear in the final \acro{EPS} graphic as a
replacement of the corresponding expression \marg{expr}. If set to
\mc{Automatic}, the internal function \mc{GuessTeX} is called, which
is essentially a wrapper for \mc{TeXForm} that adds {\$} signs around
math expressions. The options \mc{PSfragTag}, \mc{TeXPosition},
\mc{PSPosition}, \mc{PSRotation} and \mc{PSScaling} are in one-to-one
correspondence to the options of \cs{psfrag} explained in the previous
section.

As a last resort \mc{TeXShiftX/Y} provide a way to move the \LaTeX\
expression by a specified \TeX\ dimension. A disadvantage of this
method is that it does not scale with the image dimensions.

Unless the user provides specific values, \MathPSfrag\ uses the
\mc{FullGraphics} command to determine the arguments for
\mc{TeXPosition} and \mc{PSPosition}.

Unfortunately \mc{FullGraphics} does not work for three-dimensional
graphics, such that \mc{PSfragExport} falls back to
\mc{PSfragManualExport}, which does not perform any alignment
detection and can also be used for two-dimensional graphics if
automatic processing is not desired.  In these cases \mc{PSfrag} has
to be applied by hand to the argument of any \mc{Text} directive and
text producing option like \mc{PlotLabel}.  Moreover at least the
values of \mc{TeXPosition} will have to be given by the user.

Since this can be rather cumbersome for explicit tick mark specifications,
an additional convenience command has been introduced,
\begin{MyVerbatim}[commandchars=\\()]
PSfragTicks[\marg(tickspec), \oarg(options)]
\end{MyVerbatim}
which simply applies \mc{PSfrag} with the provided \oarg{options} to
any tick mark label in \marg{tickspec}. Since the tick mark
specification itself is usually quite lengthy, it is recommended to
create it with the \mc{LinTicks[\marg{from}, \marg{to}]} command from
\mc{CustomTicks} \cite{Caprio:2005}. A three-dimensional plot
illustrating this procedure is included as example code; its output is
depicted in fig.~\ref{fig:three-d}.

\begin{figure}
\centering
  \ifpdf\else\input{ex_3d-psfrag.tex}\fi  
  \mbox{\includegraphics[width=75mm]{ex_3d-psfrag.eps}}
\\[3ex]
\noindent
\begin{minipage}{0.93\linewidth}
\begin{MyVerbatim}[commandchars=\\()]
Ticks\Rule{
  PSfragTicks[LinTicks[-1,1],TeXPosition\Rule"Bc"],
  PSfragTicks[LinTicks[-1,1],TeXPosition\Rule"Bl"],
  PSfragTicks[LinTicks[-1,1],TeXPosition\Rule"Br"]}
\end{MyVerbatim}
\end{minipage}
\caption{\label{fig:three-d}Three dimensional example: As there exists
  no \mc{FullGraphics3D} command, manual labeling was preformed by
  using the above \mc{Plot3D} option.  \mc{LinTicks} is provided by
  the \mc{CustomTicks} package \cite{Caprio:2005}.}
\end{figure}

\section{UnPSfrag\label{sec:unpsfrag}}
With the commands described so far, standard \PSfrag\ \acro{EPS}
pictures can be produced. \MathPSfrag\ provides two additional
features that require a full \LaTeX\ run from within \Mathematica.
First, creation of pre-rendered \acro{EPS} images, which do not
require the \PSfrag\ package anymore. Such pre-rendered images will be
called ``unpsfraged'' henceforth.  Second, conversion to \acro{PDF}
images suitable for \PDFLaTeX\ or raster images providing a preview of
the final plot.  For both features, \MathPSfrag\ needs to know where
to find the executables for \LaTeX, \dvips\ and \Ghostscript. Unless
the binaries are in the system's execution path (as will be the case
for Unix-like operating systems), this requires the user to set the
location of the files explicitly.  The notebook containing the code
of all example plots shown in this article also contains a
comprehensive step-by-step guide of how to set up these paths
permanently. Detailed explanations can also be found in the manual.

The syntax of \mc{UnPSfrag} is
\begin{MyVerbatim}[commandchars=\\()]
UnPSfrag[{\marg(basename),\marg(epsfile),\marg(texfile)},\oarg(options)];
\end{MyVerbatim}
where \marg{basename} is used to create the names of output files by
appending suitable suffices like ``\fname{.pdf}'' or ``\fname{.eps}''.
The other two mandatory parameters should point to the tagged
\acro{EPS} and \PSfrag-\LaTeX\ files that are to be merged.
\mc{PSfragExport} returns the three file names in exactly this format
such that its output can be directly fed to \mc{UnPSfrag}.

\mc{UnPSfrag} has a number of additional options that control how
images are created. Again only the most important ones shall be given
here, while a complete list is given in the manual.

By default, \mc{UnPSfrag} only produces an \acro{EPS} and a \acro{PDF}
file, though it should be possible to generate other output formats by
setting the \mc{UnPSfragOutputFormats} option accordingly.

Furthermore \mc{UnPSfrag} shows a low resolution bitmap preview of the
final images. While this is extremely useful for checking if
\MathPSfrag\ has produced the desired results, it will enlarge the
notebook considerably, which can be inconvenient for email exchange.
Setting the \mc{PreviewDevice} option to \mc{None} will switch off
this behavior. This option is further discussed in the section
on known bugs.

When using the \mc{PSfrag} option \mc{TeXCommand} to specify \LaTeX\
code, it might require additional style files. These can be included
by \mc{UnPSfrag} when providing the appropriate \cs{usepackage}
command as a string via the \mc{TeXPreamble} option.

\mc{IncludeGraphicsOptions} is the most important option. The string it
provides is handed over to the \cs{includegraphics} command in the
\LaTeX\ file used to perform the \cs{psfrag} replacements.  The option
can therefore be used to set the size of the rendered graphics; e.g.,
by setting its value to \mc{"width=7cm"}. Note however that the final
bounding box will correctly fit the image \emph{content} instead of
exactly matching the specified \emph{size}. This mismatch is due to
the bounding box changing during \PSfrag's replacement procedure. When
an exact size is required, it can be reapplied in the manuscript,
though it is better to adjust the size from within \Mathematica, the
reason for which is explained in the next section.

\section{In the manuscript}
\mc{UnPSfrag}'ed images do not require any additional treatment or
package beyond \latex{graphics}/\latex{graphicx}. Since \acro{EPS} and
\acro{PDF} versions are created anyway, it is recommended not to
provide any suffix for the file name in the \cs{includegraphics}
command, such that the manuscript translates with both \LaTeX\ and
\PDFLaTeX. It is possible to issue the common size options, but this
will also change the labels, such that they will not have the fixed
\TeX\ sizes anymore. This can potentially reduce the fonts quality, in
particular when \TeX's old bitmapped computer modern fonts are used.
The \mc{UnPSfrag} option \mc{DvipsOptions} is by default set to
\mc{"-Ppdf"}, which makes \fname{dvips} replace bitmap fonts by
outline fonts on most systems, thus reducing the problem to a mere
mismatch between the label's size and that of the manuscript's text.
In any case it is advisable to adjust the plot size from within
\Mathematica\ before rendering by applying \mc{IncludeGraphicsOptions}
to \mc{UnPSfrag}, since rescaling at a later stage would also
affect the labels.

It is also possible to follow a ``pure'' \PSfrag\ approach, where all
replacements are performed on the level of the manuscript. This is
discussed in section \ref{sec:psfragsty}. Before, a number of usage
examples shall be given.

\section{Examples\label{sec:examples}}
We start to consider in more detail figures \ref{fig:poor} and
\ref{fig:beautiful}.  The first one has been generated using standard
\Mathematica{} commands only, for the latter, the export was carried
out with \mc{PSfragExport["example", exampleplot]//UnPSfrag}.

\begin{figure}
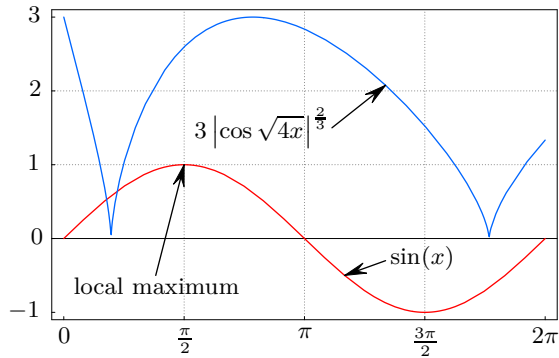

\vspace{4ex}
\centering
  \ifpdf\else\input{ex_manual-psfrag.tex}\fi  
  \mbox{\includegraphics[width=75mm]{ex_manual-psfrag.eps}}

\caption{Example plot without resorting to automatics; i.e., exported
  with \mc{PSfragManualExport}.  Additionally, the typesetting of the
  ``$\cos \dots$'' label has been manually
  improved.\vspace{1ex}\label{fig:manual}}
\end{figure}

In fig.~\ref{fig:manual} the introductory example fig.~\ref{fig:beautiful} 
was reconstructed without resorting to the
automatic positioning feature by using \mc{PSfrag} and
\mc{PSfragManualExport} only.  Additionally, one of the labels'
\LaTeX\ code was improved to be $3 \left|\cos
  \sqrt{4x}\right|$\raisebox{1ex}{$\scriptstyle\frac{2}{3}$} instead
of $3 \sqrt[3]{\cos^2(2\sqrt{x})}$.

\begin{figure}
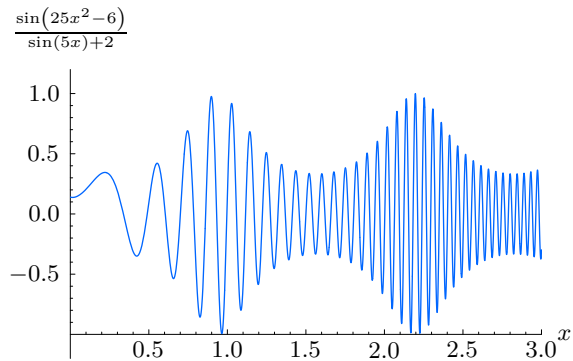

  \ifpdf\else\input{ex_hold-psfrag.tex}\fi  
  \mbox{\includegraphics[width=75mm,trim=0 0 0 -20]{ex_hold-psfrag.eps}}

\caption{\mc{HoldForm} example: Without \mc{HoldForm}, \Mathematica{}
  would normal order the label on the $y$ axis to $-\frac{\sin 6 -
    25x^2}{2+\sin(5x)}$. The \pack{CustomTicks} package has been used
  to avoid the typical stripped decimal ``1.'' on the $x$ axis.
\label{fig:hold}}
\end{figure}

Fig.~\ref{fig:hold} demonstrates compatibility with the
\pack{CustomTicks} package \cite{Caprio:2005}, which can be used to
customize tick marks, and the \mc{HoldForm} command, which can be used
to circumvent \Mathematica's automatic reordering of expressions into
a normal form.

\begin{figure}
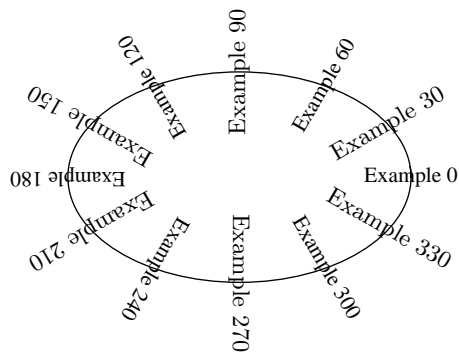

\centering
  \ifpdf\else\input{ex_rot-psfrag.tex}\fi  
  \mbox{\includegraphics[height=42.5mm]{ex_rot-psfrag.eps}}

\caption{Example for substituting rotated text. To demonstrate that
the new export function can preserve the orientation, only half of
the labels have been substituted by \LaTeX. \label{fig:rot}}
\end{figure}

While early Linux versions of \Mathematica{} do not reliably rotate
text in an interactive session, \mc{PSfragExport} has no problems in
doing so, as has been shown in fig.~\ref{fig:rot}. Note that for each
piece of text, the \mc{PSRotation} option is left unchanged
(corresponding to ``0''), thus preserving the original orientation of
the \PostScript\ text.

Finally, it has been demonstrated in fig.~\ref{fig:three-d}, that
three dimensional graphics can be processed also, even though it has
to be done manually with \mc{PSfrag} commands, since the
\mc{FullGraphics} command, which is used to extract positioning
information, only works on two dimensional graphics.

\section{PDF}
For publication of \LaTeX\ documents, the author often not only has to
provide the final manuscript but also the sources to produce them
subject to constraints set by the publisher.

Therefore, \MathPSfrag\ essentially offers two levels of image
production, a rather sophisticated ``\mc{UnPSfrag}'' approach, which
aims at producing pre-rendered stand-alone images as has been
described above, and a classical, \PSfrag\ centric approach.  In this
approach, \MathPSfrag\ is used to produce a tagged \acro{EPS} file
plus a \PSfrag\ translation file, which can convert the tags back into
readable labels during the \LaTeX\ run of the manuscript. These two
files are generated by \mc{PSfragExport}.  (Further use of
\mc{UnPSfrag} would join this pair into stand-alone graphics.)

The decision which approach to use has an impact on both the process
of image creation and on how to include the images and compile the
manuscript.  In particular, the \PSfrag\ centric approach is less
convenient and not as compatible as the \mc{UnPSfrag} approach, since
it relies on \PostScript. However it has the advantage of producing
considerably smaller images of potentially better typesetting quality
as the used fonts are guaranteed to match those of the document.
Furthermore, the font size of labels is fixed in this approach, which
is of particular importance for the bitmapped Computer Modern, though
modern \TeX\ distributions have outline fonts available, which scale
more gracefully. Still visual consistency might suffer from
differently scaled labels.

For inclusion of \acro{EPS} images into \PDFLaTeX\ manuscripts, a
number of packages exist, mainly due to the widespread use of the
equally \PostScript\ dependent \latex{pstricks} package.  Here, only
the \pstpdf\ package will be described, because it is
compatible with \pack{beamer}'s overlay techniques.  An example presentation
using \cs{psfrag} to change labels in a plot slide by slide is
included in the distribution as well as examples for the other
packages, \latex{ps4pdf} and \latex{pdftricks}, that could also be
used to achieve \PDFLaTeX\ compatibility.\footnote{I would like to
  thank Ross Moore for bringing these packages to my attention.}

\subsection{Standard PSfrag (``PostScript only'')}
The standard \PSfrag\ way is to include the \acro{EPS} and \PSfrag\
file in a manner similar to the following code.
\begin{MyVerbatim}
\begin{psfrags}
  \input{example-psfrag.tex}
  \includegraphics{example-psfrag.eps}
\end{psfrags}
\end{MyVerbatim}
The \env{psfrags} starts an empty group provided by \PSfrag{}, whose
sole purpose is making \cs{psfrag} definitions local to the following
graphic.

The produced \acro{DVI} file is then required to be converted to
\PostScript\ using \texttt{dvips}; and \acro{PDF} can only be created
by distilling from the \PostScript\ version using for example Acrobat
or \Ghostscript (\fname{ps2pdf}). Note that the popular \fname{dvipdf}
command(s) will not work.

The disadvantage of the pure \PostScript\ approach is that the
advanced typesetting features of \PDFLaTeX\ are not available.

\subsection{\PDFLaTeX\ (pst-pdf)}
\pstpdf\ extracts all images of a manuscript in an additional
\LaTeX\ run with the help of the \pack{preview} package.\footnote{A related
package is \pack{auto-pst-pdf} \cite{Robertson:2007}, which
automatizes all of the following steps provided that \PDFLaTeX\ can be
called with the \fname{-shell-escape} option.} 
These can then be turned into a \PostScript\ file and distilled to a \acro{PDF}
called \emph{image container}.  During the \PDFLaTeX\ run,
\acro{PDF} replacements for \PostScript\ parts are read from the image
container. Whenever content or order of the images are changed in the
manuscript, the image container must be regenerated.

The required steps are automated in a script accompanying
\pstpdf. Unfortunately, the predecessor package \pack{ps4pdf},
which is still installed on many systems, has a similar script sharing
the same name, \fname{ps4pdf}. It is important to ensure that the
correct version of the \fname{ps4pdf} script is used that refers to
\pstpdf.

There is one additional catch due to the way \cs{psfrag} invalidates
bounding boxes: \pstpdf\ must be loaded with the
\latex{notightpage} option and the \fname{ps4pdf} script must be called
with the \fname{-{}-crop} option.

Assuming \fname{manuscript.tex} to be the main document, either of the
following instructions produces the image container:
\begin{MyVerbatim}[commandchars=\\(),fontsize=\small]
ps4pdf --crop manuscript.tex
\textrm(or)
latex manuscript.tex
dvips -Ppdf -o manuscript-crop.ps manuscript
ps2pdf -dAutoRotatePages=/None manuscript-crop.ps \textbackslash()
\hfill()manuscript-crop.pdf
pdfcrop manuscript-crop.pdf manuscript-pics.pdf
\end{MyVerbatim}
Of course, a distiller different from \fname{ps2pdf} may be used.
Subsequently, the manuscript can be processed by \PDFLaTeX\ in the
usual manner, which will use the pre-generated pictures stored in
the \fname{-pics.pdf} file.

When uploading the manuscript to a preprint server that renames
the main manuscript file, care has to be taken to explicitely
specify the name of the image container as it will not be
found by \pstpdf\ otherwise. This can be achieved as is 
illustrated in the following example.
\begin{MyVerbatim}[fontsize=\small]
\usepackage{pst-pdf}
\renewcommand{PDFcontainer}{manuscript-pics.pdf}
\end{MyVerbatim}

The main advantage of this approach is that all images are collected in
one single file, which can be very convenient for distribution, since
it is much smaller than individual \acro{PDF} versions of each image.
This observation suggests that the collective image file is free of
double copies of resources like fonts.

\section{Known bugs and limitations\label{sec:discussion}}
\MathPSfrag{} relies on three \Mathematica{} commands: \mc{TeXForm},
\mc{FullGraphics} and \mc{AbsoluteOptions}. All are potential sources
of failure and their respective peculiarities are highly version
dependent.  A detailed analysis of potential traps is given in the
manual.  Here we shall concentrate on how to overcome certain
restrictions or problems.

For versions 4.0--5.0 \mc{TeXForm} produces \Mathematica-specific
\LaTeX\ code that requires a compatibility layer or the style files
and fonts provided by Wolfram~Inc. This is explained in more detail in
the appendix. Later \Mathematica\ versions produce \amsmath\
symbols and code, but the results will only be satisfactory for those
\Mathematica\ symbols that have a direct counterpart in \LaTeX. For
most applications this should of course suffice.

\mc{TeXForm} generates a number format that corresponds to \mc{InputForm},
while in plots \mc{TraditionalForm} would be appropriate.  It is
\emph{strongly recommended} to use the \pack{CustomTicks} package
whenever numbers on the axes are formatted unsatisfactorily. A simple 
\mc{Ticks\Rule{}LinTicks} provide as option to the \mc{Plot} command
suffices in most cases.

\mc{AbsoluteOptions} is not always faithful in the sense that
\mc{Show[myplot, AbsoluteOptions[myplot]]} will not always generate
the same plot. Most of the time, tick marks are affected. Again,
this can be avoided by employing \pack{CustomTicks}.

For three-dimensional plots, where manual alignment information has to
be provided, the \mc{PSfragTicks} command should be used for
convenience.

If the \LaTeX\ run performed by \mc{UnPSfrag} fails, error logs are
stored in internal variables and a help message will give hints of how
to identify the error.  Most of the time either a package is missing
or a malformed \LaTeX\ expression has been assigned to a label.  The
former can be solved by calling \mc{UnPSfrag} with
\mc{TeXPreamble\Rule{}"\textbackslash\textbackslash{}usepackage\{\marg{missing}\}"}
and providing the missing package(s).  The latter can be addressed by
wrapping a \mc{PSfrag} command about the offending \Mathematica\
expression and assign better \LaTeX\ code by hand using the
\mc{TeXCommand} option.

Since \cs{psfrag}-based images will in general have incorrect 
bounding boxes, \MathPSfrag\ measures the bounding box by 
converting the image to a bitmap and counting non-empty lines.
This should be more reliable than \Ghostscript's \fname{bbox} device,
which is known to sometimes produces incorrect bounding boxes.
For speed and lower memory consumption, it is however possible
to choose the \fname{bbox} device by setting the \mc{UnPSfrag}
option \mc{BoundingBoxDevice\Rule{}\{".bbox","bbox",""\}}.

In \Mathematica\ version 4.1 under \MacOSX\ (though other versions
before 5.2 may be affected), displaying of bitmap images leads to 
a kernel freeze on subsequent usage of \mc{UnPSfrag}. 
This can be avoided by setting \mc{PreviewDevice\Rule{}None} to
switch off generation of preview images.
 
\section{Conclusion}
\MathPSfrag{} provides a convenient interface to \PSfrag{} permitting
the generation of high-quality labels in \Mathematica{} graphics.
While it automatizes all tedious aspects of \PSfrag{}, it still allows
seamless overriding of all of its internal assumptions. The
possibility to create images that do not depend on \PSfrag\ anymore
provides a simple method to achieve \PDFLaTeX\ compatibility.
For convenience, a preview feature has been introduced, which 
allows the user to easily ensure that the generated labels are
both syntactically correct and yield the intended output.

Finally, \MathPSfrag{} does not provide methods to construct correct
tick mark \emph{contents} as it is strictly focused on shape. As shown
in fig.~\ref{fig:hold}, it does however integrate well with the
\pack{CustomTicks} package \cite{Caprio:2005}, which provides that
functionality.

For the future it would be interesting to incorporate the
\pack{PSfragx} extension that allows including the \cs{psfrag}
commands into the comment section of the \acro{EPS} file. It would be
certainly desirable to export the position information, which is
already extracted anyway, in the form of \cs{overpic} commands. 
In this context, the \fname{eps2pgf} \cite{PW:2007}, which can convert
\acro{EPS} files in conjunction with \PSfrag\ labels into pure
\TeX/\acro{PGF} code, is also of great interest 
as it could be easily modified to produce such \cs{overpic} commands.
This approach could allow circumvention of some of the bounding box related
difficulties arising from the \PSfrag{} approach and also achieve
\PDFLaTeX\ compatibility on a very fundamental level.
Another possibility would be a package that renders all graphics
into the correct size by calling \LaTeX\ 
from within \PDFLaTeX\ by means of the \texttt{-shell-escape}
option. 

\section*{Acknowledgments}
I am grateful to Riccardo Apreda and Robert Ei\-sen\-reich for helpful
comments and discussion. Many thanks also to Will Robertson for
extended testing and bug reports related to \MathPSfrag\ under 
version 6 of \Mathematica.

\appendix

\onecolumngrid

\section*{Program Summary}
\noindent
\begin{tabular*}{\textwidth}{@{\extracolsep\fill}lp{0.75\textwidth}}
{\em Manuscript Title:}&           \MathPSfrag: Creating publication-quality  
                                   Labels for \MathematicaR{} Plots                  \\[1.5ex]
{\em Authors:}&                    Johannes Gro{\ss}e                                \\[1.5ex]
{\em Program Title:}&              \MathPSfrag                                       \\[1.5ex]
{\em Journal Reference:}&                                                            \\[1.5ex]
{\em Catalogue identifier:}&                                                         \\[1.5ex]
{\em Program available from: }&    $ $
\url{http://wwwth.mppmu.mpg.de/members/jgrosse/mathpsfrag}\\[1.5ex]
{\em Licensing provisions:}&       CPC non-profit use license                        \\[1.5ex]
{\em Programming language:}&       \Mathematica{} 6.0, 5.2, 5.0, 4.1                 \\[1.5ex]
{\em Computer:}&                   Tested on x86 architecture                        \\[1.5ex]
{\em Operating systems:}&          Tested on Linux, Windows~XP, \MacOSX              \\[1.5ex]
{\em RAM used during test run:}&   11 Mb                                             \\[1.5ex]
{\em Keywords:}&                   Encapsulated PostScript, \Mathematica,
                                   \PSfrag, \LaTeX                                   \\[1.5ex]
{\em PACS:}&                       01.30.Rr                                          \\[1.5ex]
{\em Classification:}&             14 Graphics                                       \\[1.5ex]
{\em External routines:}&          \pack{CustomTicks} package \cite{Caprio:2005} 
                                   (recommended)                                     \\[1.5ex]
{\em Nature of problem:}&          Insufficient typesetting quality of labels in 
                                   graphics exported from \Mathematica\              \\[1.5ex]
{\em Solution method:}&            An automatic export function is provided that
                                   generates \LaTeX{} substitution labels.           \\[1.5ex]
{\em Requirements:}&               \LaTeX,  \Ghostscript, \PSfrag; recommended: 
                                   \fname{ps2eps}, Perl, \fname{pdfcrop},
                                   \LaTeX\ package \pstpdf{}                    \\[1.5ex]
{\em Restrictions:} &              The described method requires to some extent 
                                   the use of Encapsulated PostScript (\acro{EPS})
                                   graphics though conversion to \acro{PDF} is 
                                   supported. Special \Mathematica\ characters that
                                   do not have a direct counterpart in \LaTeX\ will
                                   not show satisfactory typesetting quality.                            
      
                                   For \Mathematica\ versions earlier than 5.1, the automatically 
                                   created \LaTeX\ code requires a compatibility 
                                   \LaTeX\ package, which is included in the package. 
                                   In \Mathematica\ 4.1, one of the examples does not work in its
                                   current form due to a bug in \Mathematica's 
                                   \mc{Export} command. However, the same effect 
                                   can be achieved using exclusively
                                   \MathPSfrag's rotation mechanism. Moreover, in \Mathematica\ 
                                   4.1 under \MacOSX, display of preview images 
                                   does not work.                                    \\[1.5ex]
{\em Running time:}&               1 minute for all examples in this article         \\[1.5ex]
\end{tabular*}

\section{Mathematica 4.x--5.0\label{sec:mma4}}
\MathPSfrag\ creates the \LaTeX\ code that corresponds to a particular
\Mathematica\ expression by calling \mc{TeXForm}. Starting from
version 5.1 the \mc{TeXForm} output is self-supporting code, which
only requires a few standard packages, in particular \amsmath.
Pre-5.1 \mc{TeXForm} output will require an additional style file
providing \Mathematica\ fonts that have to be configured accordingly
\cite{WRI:tetex,WRI:gsfont,WRI:intoeps}.  In the new approach, since a
large number of \Mathematica\ symbols do not have counterparts in
standard \LaTeX\ packages, some symbols are created by gluing together
existing symbols. When applied carefully, this method can achieve
acceptable results though the current implementation is not optimal
yet.

The problem with the pre-5.1 output in conjunction with the \PSfrag\
method is that it requires the \Mathematica\ fonts to be available in
the publisher's \LaTeX\ installation, while in general they will---if
at all---only be available to the author. (This should not be a
problem when the author decides to follow the \mc{UnPSfrag} approach,
where all required fonts are embedded into the image files. It is then
however necessary to include the style files by setting the according
\mc{UnPSfrag} option:
\mc{TeXPreamble\Rule{}"\textbackslash{}usepackage\{...\}"}.)

To avoid these problems in the \PSfrag\ approach requires a
compatibility package (\fname{mma4tex.tex}) that is included in the
\MathPSfrag\ distribution. Basically it parses \Mathematica\ 4.x--5.0
\LaTeX\ code and replaces it by code similar to the output of version
5.1/5.2.  When \MathPSfrag\ detects \Mathematica\ 5.0 or earlier,
\mc{PSfragExport} automatically creates \PSfrag\ files that attempt to
load \fname{mma4tex.tex} and issue a warning when unsuccessful.  There
is also a corresponding style file \fname{mma4tex.sty} that
additionally loads the required font packages and ensures that
\fname{mma4tex.tex} is only loaded once.  It is recommended to include
it in the preamble.  Therefore both files have to be installed where
\LaTeX\ can find them; e.g., in a directory \fname{tex/latex/mma4tex/}
relative to the \fname{texmf} directory of the \LaTeX\ distribution.
Thereafter the file database (``\fname{ls-R}'') has to be updated.
Again the details depend on the distribution. Typical names of the
update command are \fname{mktexlsr} or \fname{texhash}. For MiKTeX
there is the choice between the command line program (\fname{initexmf
  -u}) and a graphical configuration program (``MiKTeX Options'' or
``Settings''; press the button ``General/File Name Database/Refresh
Now'').

\bibliographystyle{elsart-num}
\bibliography{../common/mathpsfrag.bib}
\end{document}